%% file: main.tex
\documentclass[conference]{IEEEtran}
\IEEEoverridecommandlockouts
\usepackage{cite}
\usepackage{amsmath,amssymb,amsfonts}
\usepackage{graphicx}
\usepackage{siunitx}
\usepackage[caption=false,font=footnotesize]{subfig}
\DeclareSIUnit{\pairs}{pairs}
\def\BibTeX{{\rm B\kern-.05em{\sc i\kern-.025em b}\kern-.08em
    T\kern-.1667em\lower.7ex\hbox{E}\kern-.125emX}}
\begin{document}

\title{A Resource Estimation Model for the Hardware-Software Co-Design of Distributed Quantum Architectures
\thanks{This research was supported by Deakin University and CSIRO's Data61 through the Next Generation Quantum Graduate Program (118-NGQGP).}
}

\author{\IEEEauthorblockN{Raymond P. H. Wu}
\IEEEauthorblockA{\textit{School of Information Technology} \\
\textit{Deakin University}\\
Burwood, Victoria 3125, Australia \\
rayphwu@gmail.com}
\and
\IEEEauthorblockN{Chathurika Ranaweera}
\IEEEauthorblockA{\textit{School of Information Technology} \\
\textit{Deakin University}\\
Burwood, Victoria 3125, Australia \\
chathu.ranaweera@deakin.edu.au}
\and
\IEEEauthorblockN{Sutharshan Rajasegarar}
\IEEEauthorblockA{\textit{School of Information Technology} \\
\textit{Deakin University}\\
Burwood, Victoria 3125, Australia \\
sutharshan.rajasegarar@deakin.edu.au}
\and
\IEEEauthorblockN{Ria Rushin Joseph}
\IEEEauthorblockA{\textit{School of Information Technology} \\
\textit{Deakin University}\\
Burwood, Victoria 3125, Australia \\
ria.joseph@deakin.edu.au}
\and
\IEEEauthorblockN{Jinho Choi}
\IEEEauthorblockA{\textit{School of Electrical and Mechanical Engineering} \\
\textit{Adelaide University}\\
Adelaide, South Australia 5005, Australia \\
jinho.choi@adelaide.edu.au}
\and
\IEEEauthorblockN{Seng W. Loke}
\IEEEauthorblockA{\textit{School of Information Technology} \\
\textit{Deakin University}\\
Burwood, Victoria 3125, Australia \\
seng.loke@deakin.edu.au}
}

\maketitle

\begin{abstract}
    In distributed quantum computing (DQC), executing monolithic quantum circuits across multiple interconnected quantum processing units (QPUs) requires dedicated communication qubits to generate and distribute entanglement.
    Because the number of physical qubits within a QPU is finite, a trade-off emerges where allocating more communication qubits increases the capacity of quantum channels for concurrent non-local operations, but reduces the number of computational qubits available for local gate operations.
    Distributed quantum compilation routinely ignores this channel capacity, while hardware architects lack a method to determine it prior to quantum circuit partitioning.
    Moreover, scheduling entanglement on demand introduces severe latency, whereas pre-fetching exposes stored pairs to decoherence.
    We propose an economic order quantity model from perishable inventory theory to optimize the trade-off between entanglement distribution latency and the time cost of decoherence.
    The resulting estimate is driven by algorithmic demand and physical constraints, offering a dual application for the hardware-software co-design of high-performance DQC: for hardware architects, it gives the optimal allocation of dedicated communication qubits in static heterogeneous architectures; for compiler developers, it gives the optimal number to reserve dynamically in homogeneous architectures.
\end{abstract}


\section{Introduction}\label{seci}
In distributed quantum computing (DQC)\cite{loke2023,caleffi2024}, monolithic quantum circuits are partitioned into a set of sub-circuits that can be executed concurrently across multiple interconnected quantum processing units (QPUs).
Quantum communication between remote QPUs relies on teleportation protocols, which require both classical communication and the generation and distribution of entangled pairs to establish quantum channels\cite{cacciapuoti2020}.
Within each QPU, qubits are functionally classified as either computational or communication qubits.
A set of computational qubits is dedicated to executing local gate operations, while a separate set of communication qubits is reserved for generating and distributing entangled pairs across the network to enable inter-processor communication\cite{kozlowski2023}.
As the number of available QPUs in the quantum network increases, the number of possible quantum channels scales quadratically\cite{caleffi2024}: for $k$ nodes on a complete graph $K_k$, there are $k(k-1)/2$ channels, and if each channel requires $n$ communication qubits, each QPU needs $n(k-1)$.
The number of allocated communication qubits is therefore one of the constraints that limit the capacity of these quantum channels.
Because the number of physical qubits is finite, a trade-off emerges: allocating more communication qubits increases the entanglement resources for concurrent non-local operations, but reduces the number of computational qubits available for local gate operations\cite{ferrari2021}.
Optimizing this allocation is critical for the performance of DQC and represents a fundamental open problem\cite{ferrari2021}.

Distributed quantum compilation routinely ignores the capacity of quantum channels, assuming an unconstrained reservoir of communication qubits\cite{ferrari2021}.
Because the exact demand for communication qubits is only known after the qubit assignment schedule is generated from quantum circuit partitioning, compilers may produce physically infeasible schedules.
Recent works\cite{burt2024,burt2026} have acknowledged this limitation and proposed imposing the number of communication qubits as a constraint during quantum circuit partitioning, but working within tight constraints degrades solution quality and hence the performance of DQC if the architecture is designed with insufficient quantum channel capacity.
Avoiding such bottlenecks requires co-designing the architecture from the ground up.
Yet, because the communication qubit requirement is obtained only after partitioning, hardware architects and compiler developers lack a method to predetermine this optimal capacity, which is governed simultaneously by algorithmic demand, physical systems, and network configuration.

Modular architectures address quantum interconnects in two regimes\cite{awschalom2021}.
In homogeneous architectures, the roles of computational and communication qubits are dynamically reconfigurable by the compiler.
For example, trapped-ion qubits are shuttled between memory and interaction regions using dynamic electric fields in quantum charge-coupled devices (QCCDs)\cite{kielpinski2002,pino2021}, and superconducting QPUs utilize an all-to-all router reconfigured by external magnetic flux\cite{wu2024}.
In contrast, heterogeneous architectures scale by interconnecting QPUs via specialized hardware interfaces, such as trapped-ion QCCDs with photonic interconnects\cite{monroe2014} or superconducting QPUs with microwave or optical connections\cite{bravyi2022}.
These interfaces must be permanently engineered into the QPU during manufacturing, requiring hardware architects to predetermine the optimal allocation to prevent interconnect bottlenecks.

Regardless of the underlying architecture, a primary compilation challenge is scheduling entanglement generation and distribution.
An on-demand strategy generates pairs only when required, introducing severe latency that halts sub-circuit execution.
A pre-fetching strategy greedily generates and stores pairs in advance to minimize this latency, but idling exposes them to decoherence, eventually rendering them useless for computation and incurring a time cost from quantum error correction or re-execution.
Consequently, there exists an optimal number of communication qubits that minimizes the overall execution time.
Recognizing that time is an expensive resource in quantum networks\cite{kozlowski2023}, we propose an economic order quantity (EOQ) model\cite{harris1990} from perishable inventory theory\cite{nahmias1982} to optimize the trade-off between entanglement distribution latency and the time cost of decoherence.

\section{Resource Estimation Model}\label{secii}
We treat entangled pairs as perishable inventory and develop an EOQ model\cite{nahmias1982} to determine the optimal number of communication qubits $N_\text{comm}^*$ in a QPU.
The problem is formulated as optimizing the trade-off between the latency of generating and distributing entangled pairs and the time cost of decoherence while they idle.

Let $R_\text{QPU}$ be the entanglement demand rate, in pairs per unit time, for a single QPU.
For a quantum circuit with $N$ qubits and depth $D$, distributed across a quantum network with $k\geq2$ QPUs, the total number of entangled pairs $E$ required for execution follows from the qubit assignment schedule after quantum circuit partitioning.
Assuming a total execution time of $Dt_\text{gate}$, the network demand rate is $R_\text{network}=E/(Dt_\text{gate})$, where $t_\text{gate}$ is the gate time.
Since each bipartite entangled pair consumes resources at two nodes, and assuming the workload is distributed evenly across $k$ nodes, the single-QPU demand rate is given by $R_\text{QPU}=2R_\text{network}/k$.

We define $C_\text{lat}$ and $c_\text{lat}$ as the entanglement distribution time cost per batch request and per individual pair, respectively.
$C_\text{lat}$ represents the fixed, one-time network overhead for classical communication and routing setup, whereas $c_\text{lat}$ encapsulates the physical hardware execution time required to generate each successive entangled pair.
Distributing a batch of size $N_\text{comm}$ thus requires latency $C_\text{lat}+c_\text{lat}N_\text{comm}$.
We further define $C_\text{dec}$ as the effective time cost of decoherence per entangled pair per unit time.
$C_\text{dec}$ quantifies the time cost incurred when an idling entangled pair stored in a communication qubit undergoes decoherence.

Assuming continuous consumption, a pre-fetched buffer of size $N_\text{comm}$ is depleted linearly down to zero at rate $R_\text{QPU}$, so on average $N_\text{comm}/2$ communication qubits are occupied and exposed to decoherence.
The QPU queries the network at frequency $R_\text{QPU}/N_\text{comm}$ for a new batch of entangled pairs.
Although the true consumption is bursty and depends on the specific scheduling of non-local gates, we adopt the time-averaged rate $R_\text{QPU}$ as a first-order design metric: hardware allocation and pre-fetch provisioning are static decisions made prior to execution, for which the mean demand over the circuit governs the amortized cost, while instantaneous bursts are absorbed by the buffer itself.
The total system time cost is formulated as
\begin{equation}
    C(N_\text{comm})=\left(C_\text{lat}+c_\text{lat}N_\text{comm}\right)\frac{R_\text{QPU}}{N_\text{comm}}+C_\text{dec}\frac{N_\text{comm}}{2}.\label{eqii1}
\end{equation}
Setting $\mathrm{d}C(N_\text{comm})/\mathrm{d}N_\text{comm}=0$ and solving for $N_\text{comm}$ yields the unconstrained optimum
\begin{equation}
    N_\text{comm}=\sqrt{\frac{2C_\text{lat}R_\text{QPU}}{C_\text{dec}}},\label{eqii2}
\end{equation}
analogous to the result of the EOQ model\cite{harris1990,nahmias1982}.
This gives an architectural insight: the theoretical optimal capacity of quantum channels is independent of the per-pair cost $c_\text{lat}$ and is driven solely by the demand rate $R_\text{QPU}$, batch cost $C_\text{lat}$, and decoherence cost $C_\text{dec}$.

Assuming Markovian noise dominated by pure dephasing, we model the fidelity of an idling state as $F(t)=e^{-t/T_2}$\cite{nielsen2010}, where $T_2$ is the coherence time.
Since error suppression in quantum error correction requires a physical error rate below a critical threshold\cite{google2025}, maintaining the fidelity above an algorithmic requirement $F_\text{th}$ bounds the idling time by $t_\text{max}=-T_2\ln F_\text{th}$.
A buffer of size $N_\text{comm}$ depleted at rate $R_\text{QPU}$ has maximum storage duration $N_\text{comm}/R_\text{QPU}$, imposing $N_\text{comm}/R_\text{QPU}\leq t_\text{max}$.
Bounding \eqref{eqii2} by this usable lifetime, the optimal number of communication qubits is the piecewise minimum
\begin{equation}
    N_\text{comm}^*=\min\left(\sqrt{\frac{2C_\text{lat}R_\text{QPU}}{C_\text{dec}}},-R_\text{QPU}T_2\ln F_\text{th}\right).\label{eqii3}
\end{equation}
This estimate bridges hardware physical constraints and software algorithmic demand: for static heterogeneous architectures, it gives the optimal allocation of dedicated communication qubits, and for dynamic homogeneous architectures, it gives the optimal number for compilers to reserve, with the reserved qubits returned to computational roles upon completion.

\subsection{Cutoff Strategy}\label{seciia}
Consuming a decohered pair introduces uncorrectable errors, typically necessitating re-execution of the shot.
To avoid this catastrophic time penalty, we assume the compiler employs a proactive cutoff strategy\cite{li2021} that tracks the idle duration of pre-fetched pairs and discards any exceeding $t_\text{max}=-T_2\ln F_\text{th}$.
On discarding an expired pair, the QPU halts and requests a replacement, incurring a penalty dominated by the fixed latency $C_\text{lat}$.
Amortizing $C_\text{lat}$ over the usable lifetime yields an effective decoherence penalty rate
\begin{equation}
    C_\text{dec}=\frac{C_\text{lat}}{-T_2\ln F_\text{th}}.\label{eqiia1}
\end{equation}
Substituting \eqref{eqiia1} into \eqref{eqii3} gives
\begin{equation}
    N_\text{comm}^*=\min\left(\sqrt{-2R_\text{QPU}T_2\ln F_\text{th}},-R_\text{QPU}T_2\ln F_\text{th}\right).\label{eqiia2}
\end{equation}
The optimum then depends entirely on the algorithmic demand $R_\text{QPU}$, the fidelity threshold $F_\text{th}$, and the coherence time $T_2$, and not on $C_\text{lat}$: the local QPU architecture does not depend on the global quantum network.
This indicates modular decoupling in distributed quantum architectures.

\section{Results}\label{seciii}
Table~\ref{tab1} lists representative order-of-magnitude parameters used to evaluate the model.
The entanglement distribution latency depends on the network scale: for a quantum local area network (LAN, \qty[parse-numbers=false]{\lesssim10}{\meter}) it is dominated by hardware overhead, while for a quantum wide area network (WAN, \qty[parse-numbers=false]{\gtrsim100}{\kilo\meter}) it is dominated by the photon travel time in optical fiber.

\begin{table}[htbp]
    \renewcommand{\arraystretch}{1.3}
    \caption{Representative Parameters for Distributed Quantum Architectures}
    \begin{center}
        \begin{tabular}{llc}
            \hline
            \textbf{Parameter}&\textbf{Technology}&\textbf{Value}\\
            \hline
            Gate time $t_\text{gate}$&Superconducting (transmon)&\qtyrange[range-phrase=--]{\sim10}{100}{\nano\second}\cite{gambetta2017}\\
            Coherence $T_2$&Superconducting (transmon)&\qty{\sim100}{\micro\second}\cite{gambetta2017,siddiqi2021}\\
            Threshold $F_\text{th}$&Surface code&\qtyrange[range-phrase=--]{98.6}{98.9}{\percent}\cite{wang2011}\\
            Distribution $C_\text{lat}$&Quantum LAN (\qty[parse-numbers=false]{\lesssim10}{\meter})&\qtyrange[range-phrase=--]{0.18}{6}{\micro\second}\cite{kurpiers2018,axline2018}\\
            &Quantum WAN (\qty[parse-numbers=false]{\gtrsim100}{\kilo\meter})&\qty{\sim1}{\milli\second}\cite{azuma2023}\\
            \hline
        \end{tabular}
        \label{tab1}
    \end{center}
\end{table}

\subsection{Superconducting Transmon Qubits}\label{seciiia}
\begin{figure}[!t]
    \centering
    \subfloat[Time cost as a function of the number of communication qubits $N_\text{comm}$, illustrating the trade-off between latency and decoherence.]{\resizebox{\columnwidth}{!}{\input{fig1a.tex}}
    \label{fig1a}}
    \hfil
    \subfloat[Optimal number of communication qubits $N_\text{comm}^*$ against fidelity threshold $F_\text{th}$ for different entanglement demand rates $R_\text{QPU}$.]{\resizebox{\columnwidth}{!}{\input{fig1b.tex}}
    \label{fig1b}}
    \caption{Estimation of the optimal number of communication qubits for distributed quantum architectures.}
    \label{fig1}
\end{figure}
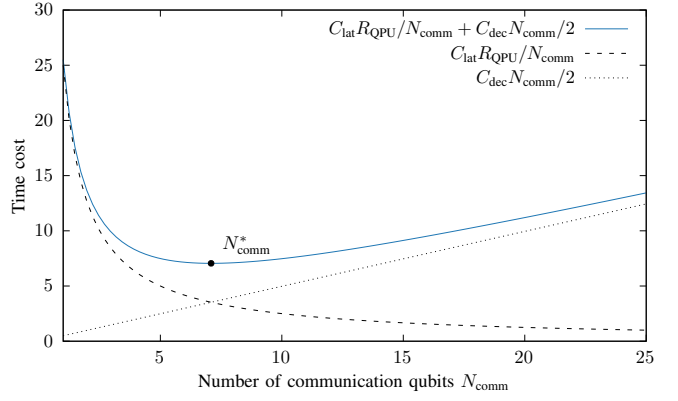
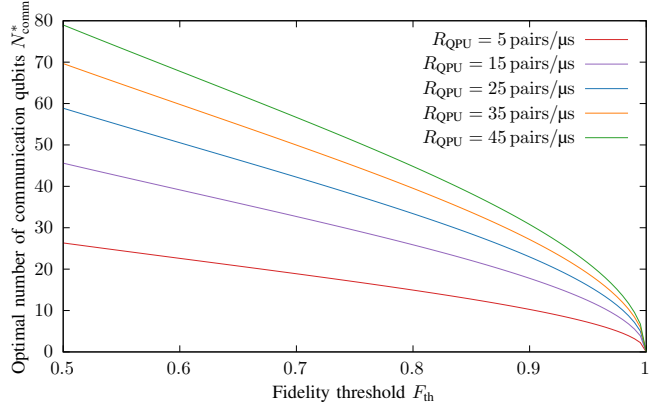

We work through the case of superconducting transmon qubits, for which $t_\text{gate}\sim\qty{100}{\nano\second}$\cite{gambetta2017} and $T_2\sim\qty{100}{\micro\second}$\cite{gambetta2017,siddiqi2021}.
The entanglement demand rate $R_\text{QPU}$ depends on quantum circuit partitioning.
For a quantum circuit with $N$ qubits and depth $D$ distributed across $k\geq2$ QPUs, the worst case requires $D\lfloor N/2\rfloor$ pairs.
Reference~\cite{burt2026} reports an entangled pair fraction (the ratio of required pairs to the worst case $D\lfloor N/2\rfloor$) of $0.023$ for QASM-large circuits\cite{li2023} partitioned over $\numlist{2;3;4}$ QPUs.
Taking a conservative fraction of $0.1$ for $N=100$ and $D=1000$, distributed across $k=4$ QPUs, the worst-case $1000\lfloor100/2\rfloor=50000$ two-qubit gates give $R_\text{network}=(\numproduct{50000x0.1})/(1000\times\qty{100}{\nano\second})=\qty[per-mode=symbol]{50}{\pairs\per\micro\second}$, hence $R_\text{QPU}=\qty[per-mode=symbol]{25}{\pairs\per\micro\second}$.
With a surface-code target of $F_\text{th}=0.99$\cite{wang2011}, \eqref{eqiia2} gives $N_\text{comm}^*=7.1$.
A QPU with $N_\text{phys}=32$ physical qubits then supports this computation with $N_\text{comp}=25$ computational and $N_\text{comm}=7$ communication qubits, dedicating $\qty{22}{\percent}$ of quantum resources to communication.
For a quantum LAN with $C_\text{lat}=\qty{1}{\micro\second}$, the cost, plotted in Fig.~\ref{fig1a}, is minimized at $N_\text{comm}=7.1$.
The effect of $F_\text{th}$ for different values of $R_\text{QPU}$ is shown in Fig.~\ref{fig1b}.

\subsection{Suboptimal Allocation Penalties}\label{seciiib}
Here, we address the possibility that the allocation of communication qubits is suboptimal due to hardware constraints.
Let $N_\text{phys}$ be the number of physical qubits and $N_\text{comp}$ be the number of computational qubits on a single QPU.
The maximum possible allocation of communication qubits is $N_{\text{comm},\text{max}}=N_\text{phys}-N_\text{comp}$.
If $N_\text{comm}^*\leq N_{\text{comm},\text{max}}$, the compiler allocates $N_\text{comm}^*$ communication qubits and the quantum circuit executes at the theoretical optimum.
However, if $N_\text{comm}^*>N_{\text{comm},\text{max}}$, the compiler is forced to limit the allocation to $N_\text{comm}=N_{\text{comm},\text{max}}$ and the system incurs a time penalty.
A quantifiable estimate of this penalty is given by the EOQ model as the difference in the total system time cost $C(N_{\text{comm},\text{max}})-C(N_\text{comm}^*)$.

\section{Challenges and Outlook}\label{seciv}
Several extensions are needed before deployment.
The demand rate $R_\text{QPU}$ depends on the targeted quantum algorithm and partitioning quality, so benchmarking diverse quantum circuits and partitioning algorithms is required for accurate calibration, together with more experimental data on entanglement distribution latency.
Because realistic demand is bursty and stochastic rather than continuous, extending this model with probabilistic error modeling and stochastic demand is a key step toward a dynamic compiler module.
Finally, the optimal allocation is further affected by entanglement management strategies: the consumption order of stored pairs (first-in-first-out versus last-in-first-out), and entanglement purification\cite{bennett1996}, which consumes multiple low-fidelity pairs to yield fewer high-fidelity ones.

\section{Conclusion}\label{secv}
We proposed a resource estimation model that resolves the disconnect between quantum hardware design and compilation, optimizing the capacity of quantum channels by minimizing the total time cost of entanglement distribution and decoherence.
Applied to superconducting transmon qubits with realistic parameters, the model yields an estimate driven by algorithmic demand and physical constraints.
It offers a dual application for the hardware-software co-design of high-performance DQC: the optimal allocation of dedicated communication qubits for hardware architects designing static heterogeneous architectures, and the optimal number for compilers to reserve dynamically in homogeneous architectures.

\bibliographystyle{IEEEtran}
\bibliography{IEEEabrv,ref}

\end{document}

%% file: fig1a.tex
\begingroup
  \makeatletter
  \providecommand\color[2][]{%
    \GenericError{(gnuplot) \space\space\space\@spaces}{%
      Package color not loaded in conjunction with
      terminal option `colourtext'%
    }{See the gnuplot documentation for explanation.%
    }{Either use 'blacktext' in gnuplot or load the package
      color.sty in LaTeX.}%
    \renewcommand\color[2][]{}%
  }%
  \providecommand\includegraphics[2][]{%
    \GenericError{(gnuplot) \space\space\space\@spaces}{%
      Package graphicx or graphics not loaded%
    }{See the gnuplot documentation for explanation.%
    }{The gnuplot epslatex terminal needs graphicx.sty or graphics.sty.}%
    \renewcommand\includegraphics[2][]{}%
  }%
  \providecommand\rotatebox[2]{#2}%
  \@ifundefined{ifGPcolor}{%
    \newif\ifGPcolor
    \GPcolortrue
  }{}%
  \@ifundefined{ifGPblacktext}{%
    \newif\ifGPblacktext
    \GPblacktexttrue
  }{}%
  \let\gplgaddtomacro\g@addto@macro
  \gdef\gplbacktext{}%
  \gdef\gplfronttext{}%
  \makeatother
  \ifGPblacktext
    \def\colorrgb#1{}%
    \def\colorgray#1{}%
  \else
    \ifGPcolor
      \def\colorrgb#1{\color[rgb]{#1}}%
      \def\colorgray#1{\color[gray]{#1}}%
      \expandafter\def\csname LTw\endcsname{\color{white}}%
      \expandafter\def\csname LTb\endcsname{\color{black}}%
      \expandafter\def\csname LTa\endcsname{\color{black}}%
      \expandafter\def\csname LT0\endcsname{\color[rgb]{1,0,0}}%
      \expandafter\def\csname LT1\endcsname{\color[rgb]{0,1,0}}%
      \expandafter\def\csname LT2\endcsname{\color[rgb]{0,0,1}}%
      \expandafter\def\csname LT3\endcsname{\color[rgb]{1,0,1}}%
      \expandafter\def\csname LT4\endcsname{\color[rgb]{0,1,1}}%
      \expandafter\def\csname LT5\endcsname{\color[rgb]{1,1,0}}%
      \expandafter\def\csname LT6\endcsname{\color[rgb]{0,0,0}}%
      \expandafter\def\csname LT7\endcsname{\color[rgb]{1,0.3,0}}%
      \expandafter\def\csname LT8\endcsname{\color[rgb]{0.5,0.5,0.5}}%
    \else
      \def\colorrgb#1{\color{black}}%
      \def\colorgray#1{\color[gray]{#1}}%
      \expandafter\def\csname LTw\endcsname{\color{white}}%
      \expandafter\def\csname LTb\endcsname{\color{black}}%
      \expandafter\def\csname LTa\endcsname{\color{black}}%
      \expandafter\def\csname LT0\endcsname{\color{black}}%
      \expandafter\def\csname LT1\endcsname{\color{black}}%
      \expandafter\def\csname LT2\endcsname{\color{black}}%
      \expandafter\def\csname LT3\endcsname{\color{black}}%
      \expandafter\def\csname LT4\endcsname{\color{black}}%
      \expandafter\def\csname LT5\endcsname{\color{black}}%
      \expandafter\def\csname LT6\endcsname{\color{black}}%
      \expandafter\def\csname LT7\endcsname{\color{black}}%
      \expandafter\def\csname LT8\endcsname{\color{black}}%
    \fi
  \fi
    \setlength{\unitlength}{0.0500bp}%
    \ifx\gptboxheight\undefined%
      \newlength{\gptboxheight}%
      \newlength{\gptboxwidth}%
      \newsavebox{\gptboxtext}%
    \fi%
    \setlength{\fboxrule}{0.5pt}%
    \setlength{\fboxsep}{1pt}%
    \definecolor{tbcol}{rgb}{1,1,1}%
\begin{picture}(7200.00,4320.00)%
    \gplgaddtomacro\gplbacktext{%
      \csname LTb\endcsname
      \put(518,562){\makebox(0,0)[r]{\strut{}$0$}}%
      \csname LTb\endcsname
      \put(518,1156){\makebox(0,0)[r]{\strut{}$5$}}%
      \csname LTb\endcsname
      \put(518,1749){\makebox(0,0)[r]{\strut{}$10$}}%
      \csname LTb\endcsname
      \put(518,2343){\makebox(0,0)[r]{\strut{}$15$}}%
      \csname LTb\endcsname
      \put(518,2936){\makebox(0,0)[r]{\strut{}$20$}}%
      \csname LTb\endcsname
      \put(518,3530){\makebox(0,0)[r]{\strut{}$25$}}%
      \csname LTb\endcsname
      \put(518,4124){\makebox(0,0)[r]{\strut{}$30$}}%
      \csname LTb\endcsname
      \put(1661,386){\makebox(0,0){\strut{}$5$}}%
      \csname LTb\endcsname
      \put(2967,386){\makebox(0,0){\strut{}$10$}}%
      \csname LTb\endcsname
      \put(4273,386){\makebox(0,0){\strut{}$15$}}%
      \csname LTb\endcsname
      \put(5580,386){\makebox(0,0){\strut{}$20$}}%
      \csname LTb\endcsname
      \put(6886,386){\makebox(0,0){\strut{}$25$}}%
    }%
    \gplgaddtomacro\gplfronttext{%
      \csname LTb\endcsname
      \put(6130,3921){\makebox(0,0)[r]{\strut{}$C_\text{lat}R_\text{QPU}/N_\text{comm}+C_\text{dec}N_\text{comm}/2$}}%
      \csname LTb\endcsname
      \put(6130,3658){\makebox(0,0)[r]{\strut{}$C_\text{lat}R_\text{QPU}/N_\text{comm}$}}%
      \csname LTb\endcsname
      \put(6130,3394){\makebox(0,0)[r]{\strut{}$C_\text{dec}N_\text{comm}/2$}}%
      \csname LTb\endcsname
      \put(161,2343){\rotatebox{-270.00}{\makebox(0,0){\strut{}Time cost}}}%
      \csname LTb\endcsname
      \put(3751,123){\makebox(0,0){\strut{}Number of communication qubits $N_\text{comm}$}}%
      \csname LTb\endcsname
      \put(2340,1611){\makebox(0,0)[l]{\strut{}$N_\text{comm}^*$}}%
    }%
    \gplbacktext
    \put(0,0){\includegraphics[width={360.00bp},height={216.00bp}]{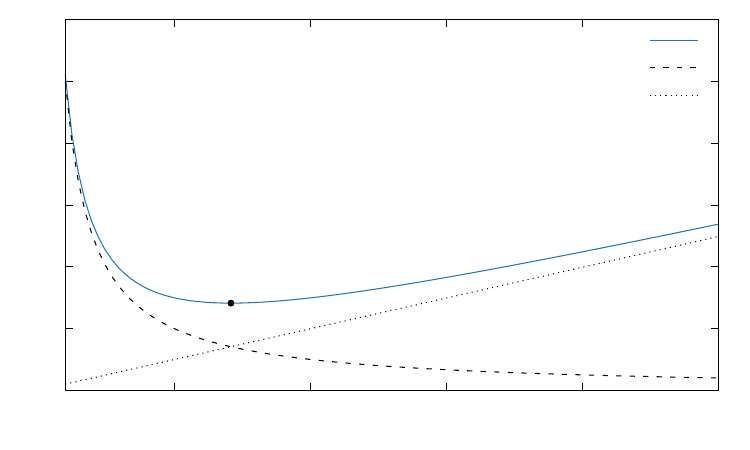}}%
    \gplfronttext
  \end{picture}%
\endgroup

%% file: fig1b.tex
\begingroup
  \makeatletter
  \providecommand\color[2][]{%
    \GenericError{(gnuplot) \space\space\space\@spaces}{%
      Package color not loaded in conjunction with
      terminal option `colourtext'%
    }{See the gnuplot documentation for explanation.%
    }{Either use 'blacktext' in gnuplot or load the package
      color.sty in LaTeX.}%
    \renewcommand\color[2][]{}%
  }%
  \providecommand\includegraphics[2][]{%
    \GenericError{(gnuplot) \space\space\space\@spaces}{%
      Package graphicx or graphics not loaded%
    }{See the gnuplot documentation for explanation.%
    }{The gnuplot epslatex terminal needs graphicx.sty or graphics.sty.}%
    \renewcommand\includegraphics[2][]{}%
  }%
  \providecommand\rotatebox[2]{#2}%
  \@ifundefined{ifGPcolor}{%
    \newif\ifGPcolor
    \GPcolortrue
  }{}%
  \@ifundefined{ifGPblacktext}{%
    \newif\ifGPblacktext
    \GPblacktexttrue
  }{}%
  \let\gplgaddtomacro\g@addto@macro
  \gdef\gplbacktext{}%
  \gdef\gplfronttext{}%
  \makeatother
  \ifGPblacktext
    \def\colorrgb#1{}%
    \def\colorgray#1{}%
  \else
    \ifGPcolor
      \def\colorrgb#1{\color[rgb]{#1}}%
      \def\colorgray#1{\color[gray]{#1}}%
      \expandafter\def\csname LTw\endcsname{\color{white}}%
      \expandafter\def\csname LTb\endcsname{\color{black}}%
      \expandafter\def\csname LTa\endcsname{\color{black}}%
      \expandafter\def\csname LT0\endcsname{\color[rgb]{1,0,0}}%
      \expandafter\def\csname LT1\endcsname{\color[rgb]{0,1,0}}%
      \expandafter\def\csname LT2\endcsname{\color[rgb]{0,0,1}}%
      \expandafter\def\csname LT3\endcsname{\color[rgb]{1,0,1}}%
      \expandafter\def\csname LT4\endcsname{\color[rgb]{0,1,1}}%
      \expandafter\def\csname LT5\endcsname{\color[rgb]{1,1,0}}%
      \expandafter\def\csname LT6\endcsname{\color[rgb]{0,0,0}}%
      \expandafter\def\csname LT7\endcsname{\color[rgb]{1,0.3,0}}%
      \expandafter\def\csname LT8\endcsname{\color[rgb]{0.5,0.5,0.5}}%
    \else
      \def\colorrgb#1{\color{black}}%
      \def\colorgray#1{\color[gray]{#1}}%
      \expandafter\def\csname LTw\endcsname{\color{white}}%
      \expandafter\def\csname LTb\endcsname{\color{black}}%
      \expandafter\def\csname LTa\endcsname{\color{black}}%
      \expandafter\def\csname LT0\endcsname{\color{black}}%
      \expandafter\def\csname LT1\endcsname{\color{black}}%
      \expandafter\def\csname LT2\endcsname{\color{black}}%
      \expandafter\def\csname LT3\endcsname{\color{black}}%
      \expandafter\def\csname LT4\endcsname{\color{black}}%
      \expandafter\def\csname LT5\endcsname{\color{black}}%
      \expandafter\def\csname LT6\endcsname{\color{black}}%
      \expandafter\def\csname LT7\endcsname{\color{black}}%
      \expandafter\def\csname LT8\endcsname{\color{black}}%
    \fi
  \fi
    \setlength{\unitlength}{0.0500bp}%
    \ifx\gptboxheight\undefined%
      \newlength{\gptboxheight}%
      \newlength{\gptboxwidth}%
      \newsavebox{\gptboxtext}%
    \fi%
    \setlength{\fboxrule}{0.5pt}%
    \setlength{\fboxsep}{1pt}%
    \definecolor{tbcol}{rgb}{1,1,1}%
\begin{picture}(7200.00,4320.00)%
    \gplgaddtomacro\gplbacktext{%
      \csname LTb\endcsname
      \put(518,562){\makebox(0,0)[r]{\strut{}$0$}}%
      \csname LTb\endcsname
      \put(518,1008){\makebox(0,0)[r]{\strut{}$10$}}%
      \csname LTb\endcsname
      \put(518,1453){\makebox(0,0)[r]{\strut{}$20$}}%
      \csname LTb\endcsname
      \put(518,1898){\makebox(0,0)[r]{\strut{}$30$}}%
      \csname LTb\endcsname
      \put(518,2343){\makebox(0,0)[r]{\strut{}$40$}}%
      \csname LTb\endcsname
      \put(518,2788){\makebox(0,0)[r]{\strut{}$50$}}%
      \csname LTb\endcsname
      \put(518,3233){\makebox(0,0)[r]{\strut{}$60$}}%
      \csname LTb\endcsname
      \put(518,3678){\makebox(0,0)[r]{\strut{}$70$}}%
      \csname LTb\endcsname
      \put(518,4124){\makebox(0,0)[r]{\strut{}$80$}}%
      \csname LTb\endcsname
      \put(616,386){\makebox(0,0){\strut{}$0.5$}}%
      \csname LTb\endcsname
      \put(1870,386){\makebox(0,0){\strut{}$0.6$}}%
      \csname LTb\endcsname
      \put(3124,386){\makebox(0,0){\strut{}$0.7$}}%
      \csname LTb\endcsname
      \put(4378,386){\makebox(0,0){\strut{}$0.8$}}%
      \csname LTb\endcsname
      \put(5632,386){\makebox(0,0){\strut{}$0.9$}}%
      \csname LTb\endcsname
      \put(6886,386){\makebox(0,0){\strut{}$1$}}%
    }%
    \gplgaddtomacro\gplfronttext{%
      \csname LTb\endcsname
      \put(6130,3921){\makebox(0,0)[r]{\strut{}$R_\text{QPU}=\qty[per-mode=symbol]{5}{\pairs\per\micro\second}$}}%
      \csname LTb\endcsname
      \put(6130,3658){\makebox(0,0)[r]{\strut{}$R_\text{QPU}=\qty[per-mode=symbol]{15}{\pairs\per\micro\second}$}}%
      \csname LTb\endcsname
      \put(6130,3394){\makebox(0,0)[r]{\strut{}$R_\text{QPU}=\qty[per-mode=symbol]{25}{\pairs\per\micro\second}$}}%
      \csname LTb\endcsname
      \put(6130,3130){\makebox(0,0)[r]{\strut{}$R_\text{QPU}=\qty[per-mode=symbol]{35}{\pairs\per\micro\second}$}}%
      \csname LTb\endcsname
      \put(6130,2866){\makebox(0,0)[r]{\strut{}$R_\text{QPU}=\qty[per-mode=symbol]{45}{\pairs\per\micro\second}$}}%
      \csname LTb\endcsname
      \put(161,2343){\rotatebox{-270.00}{\makebox(0,0){\strut{}Optimal number of communication qubits $N_\text{comm}^*$}}}%
      \csname LTb\endcsname
      \put(3751,123){\makebox(0,0){\strut{}Fidelity threshold $F_\text{th}$}}%
    }%
    \gplbacktext
    \put(0,0){\includegraphics[width={360.00bp},height={216.00bp}]{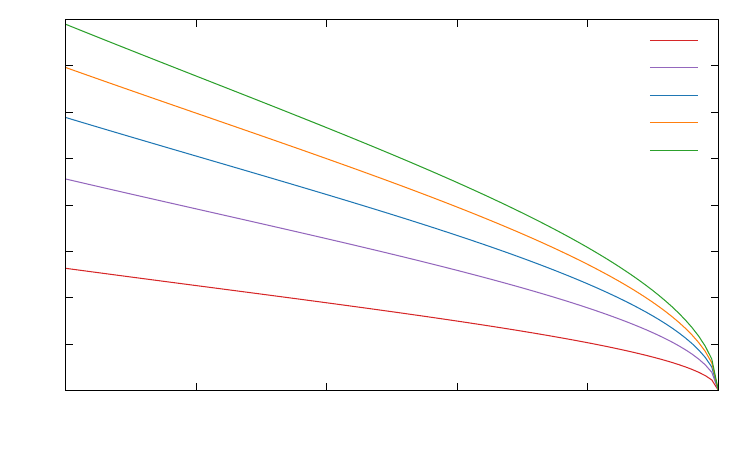}}%
    \gplfronttext
  \end{picture}%
\endgroup

%% file: main.bbl
\begin{thebibliography}{10}
\providecommand{\url}[1]{#1}
\csname url@samestyle\endcsname
\providecommand{\newblock}{\relax}
\providecommand{\bibinfo}[2]{#2}
\providecommand{\BIBentrySTDinterwordspacing}{\spaceskip=0pt\relax}
\providecommand{\BIBentryALTinterwordstretchfactor}{4}
\providecommand{\BIBentryALTinterwordspacing}{\spaceskip=\fontdimen2\font plus
\BIBentryALTinterwordstretchfactor\fontdimen3\font minus \fontdimen4\font\relax}
\providecommand{\BIBforeignlanguage}[2]{{%
\expandafter\ifx\csname l@#1\endcsname\relax
\typeout{** WARNING: IEEEtran.bst: No hyphenation pattern has been}%
\typeout{** loaded for the language `#1'. Using the pattern for}%
\typeout{** the default language instead.}%
\else
\language=\csname l@#1\endcsname
\fi
#2}}
\providecommand{\BIBdecl}{\relax}
\BIBdecl

\bibitem{loke2023}
S.~W. Loke, \emph{From distributed quantum computing to quantum internet computing: an introduction}.\hskip 1em plus 0.5em minus 0.4em\relax John Wiley \& Sons, 2023.

\bibitem{caleffi2024}
\BIBentryALTinterwordspacing
M.~Caleffi, M.~Amoretti, D.~Ferrari, J.~Illiano, A.~Manzalini, and A.~S. Cacciapuoti, ``Distributed quantum computing: A survey,'' \emph{Computer Networks}, vol. 254, p. 110672, 2024. [Online]. Available: \url{https://www.sciencedirect.com/science/article/pii/S1389128624005048}
\BIBentrySTDinterwordspacing

\bibitem{cacciapuoti2020}
A.~S. Cacciapuoti, M.~Caleffi, R.~Van~Meter, and L.~Hanzo, ``When entanglement meets classical communications: Quantum teleportation for the quantum internet,'' \emph{IEEE Transactions on Communications}, vol.~68, no.~6, pp. 3808--3833, 2020.

\bibitem{kozlowski2023}
\BIBentryALTinterwordspacing
W.~Kozlowski, S.~Wehner, R.~V. Meter, B.~Rijsman, A.~S. Cacciapuoti, M.~Caleffi, and S.~Nagayama, ``{Architectural Principles for a Quantum Internet},'' RFC 9340, Mar. 2023. [Online]. Available: \url{https://www.rfc-editor.org/info/rfc9340}
\BIBentrySTDinterwordspacing

\bibitem{ferrari2021}
D.~Ferrari, A.~S. Cacciapuoti, M.~Amoretti, and M.~Caleffi, ``Compiler design for distributed quantum computing,'' \emph{IEEE Transactions on Quantum Engineering}, vol.~2, pp. 1--20, 2021.

\bibitem{burt2024}
F.~Burt, K.-C. Chen, and K.~K. Leung, ``Generalised circuit partitioning for distributed quantum computing,'' in \emph{2024 IEEE International Conference on Quantum Computing and Engineering (QCE)}, vol.~02, 2024, pp. 173--178.

\bibitem{burt2026}
\BIBentryALTinterwordspacing
------, ``A {M}ultilevel {F}ramework for {P}artitioning {Q}uantum {C}ircuits,'' \emph{{Quantum}}, vol.~10, p. 1984, Jan. 2026. [Online]. Available: \url{https://doi.org/10.22331/q-2026-01-22-1984}
\BIBentrySTDinterwordspacing

\bibitem{awschalom2021}
\BIBentryALTinterwordspacing
D.~Awschalom, K.~K. Berggren, H.~Bernien, S.~Bhave, L.~D. Carr, P.~Davids, S.~E. Economou, D.~Englund, A.~Faraon, M.~Fejer, S.~Guha, M.~V. Gustafsson, E.~Hu, L.~Jiang, J.~Kim, B.~Korzh, P.~Kumar, P.~G. Kwiat, M.~Lon\ifmmode~\check{c}\else \v{c}\fi{}ar, M.~D. Lukin, D.~A. Miller, C.~Monroe, S.~W. Nam, P.~Narang, J.~S. Orcutt, M.~G. Raymer, A.~H. Safavi-Naeini, M.~Spiropulu, K.~Srinivasan, S.~Sun, J.~Vu\ifmmode \check{c}\else \v{c}\fi{}kovi\ifmmode~\acute{c}\else \'{c}\fi{}, E.~Waks, R.~Walsworth, A.~M. Weiner, and Z.~Zhang, ``Development of quantum interconnects ({Q}u{IC}s) for next-generation information technologies,'' \emph{PRX Quantum}, vol.~2, p. 017002, Feb 2021. [Online]. Available: \url{https://link.aps.org/doi/10.1103/PRXQuantum.2.017002}
\BIBentrySTDinterwordspacing

\bibitem{kielpinski2002}
D.~Kielpinski, C.~Monroe, and D.~J. Wineland, ``Architecture for a large-scale ion-trap quantum computer,'' \emph{Nature}, vol. 417, no. 6890, pp. 709--711, 2002.

\bibitem{pino2021}
J.~M. Pino, J.~M. Dreiling, C.~Figgatt, J.~P. Gaebler, S.~A. Moses, M.~Allman, C.~Baldwin, M.~Foss-Feig, D.~Hayes, K.~Mayer \emph{et~al.}, ``Demonstration of the trapped-ion quantum ccd computer architecture,'' \emph{Nature}, vol. 592, no. 7853, pp. 209--213, 2021.

\bibitem{wu2024}
\BIBentryALTinterwordspacing
X.~Wu, H.~Yan, G.~Andersson, A.~Anferov, M.-H. Chou, C.~R. Conner, J.~Grebel, Y.~J. Joshi, S.~Li, J.~M. Miller, R.~G. Povey, H.~Qiao, and A.~N. Cleland, ``Modular quantum processor with an all-to-all reconfigurable router,'' \emph{Phys. Rev. X}, vol.~14, p. 041030, Nov 2024. [Online]. Available: \url{https://link.aps.org/doi/10.1103/PhysRevX.14.041030}
\BIBentrySTDinterwordspacing

\bibitem{monroe2014}
\BIBentryALTinterwordspacing
C.~Monroe, R.~Raussendorf, A.~Ruthven, K.~R. Brown, P.~Maunz, L.-M. Duan, and J.~Kim, ``Large-scale modular quantum-computer architecture with atomic memory and photonic interconnects,'' \emph{Phys. Rev. A}, vol.~89, p. 022317, Feb 2014. [Online]. Available: \url{https://link.aps.org/doi/10.1103/PhysRevA.89.022317}
\BIBentrySTDinterwordspacing

\bibitem{bravyi2022}
\BIBentryALTinterwordspacing
S.~Bravyi, O.~Dial, J.~M. Gambetta, D.~Gil, and Z.~Nazario, ``The future of quantum computing with superconducting qubits,'' \emph{Journal of Applied Physics}, vol. 132, no.~16, p. 160902, 10 2022. [Online]. Available: \url{https://doi.org/10.1063/5.0082975}
\BIBentrySTDinterwordspacing

\bibitem{harris1990}
\BIBentryALTinterwordspacing
F.~W. Harris, ``How many parts to make at once,'' \emph{Operations Research}, vol.~38, no.~6, pp. 947--950, 1990. [Online]. Available: \url{https://doi.org/10.1287/opre.38.6.947}
\BIBentrySTDinterwordspacing

\bibitem{nahmias1982}
\BIBentryALTinterwordspacing
S.~Nahmias, ``Perishable inventory theory: A review,'' \emph{Operations Research}, vol.~30, no.~4, pp. 680--708, 1982. [Online]. Available: \url{https://doi.org/10.1287/opre.30.4.680}
\BIBentrySTDinterwordspacing

\bibitem{nielsen2010}
M.~A. Nielsen and I.~L. Chuang, \emph{Quantum Computation and Quantum Information: 10th Anniversary Edition}.\hskip 1em plus 0.5em minus 0.4em\relax Cambridge University Press, 2010.

\bibitem{google2025}
{Google Quantum AI and Collaborators}, ``Quantum error correction below the surface code threshold,'' \emph{Nature}, vol. 638, no. 8052, pp. 920--926, 2025.

\bibitem{li2021}
B.~Li, T.~Coopmans, and D.~Elkouss, ``Efficient optimization of cutoffs in quantum repeater chains,'' \emph{IEEE Transactions on Quantum Engineering}, vol.~2, pp. 1--15, 2021.

\bibitem{gambetta2017}
J.~M. Gambetta, J.~M. Chow, and M.~Steffen, ``Building logical qubits in a superconducting quantum computing system,'' \emph{npj quantum information}, vol.~3, no.~1, p.~2, 2017.

\bibitem{siddiqi2021}
I.~Siddiqi, ``Engineering high-coherence superconducting qubits,'' \emph{Nature Reviews Materials}, vol.~6, no.~10, pp. 875--891, 2021.

\bibitem{wang2011}
\BIBentryALTinterwordspacing
D.~S. Wang, A.~G. Fowler, and L.~C.~L. Hollenberg, ``Surface code quantum computing with error rates over 1\%,'' \emph{Phys. Rev. A}, vol.~83, p. 020302, Feb 2011. [Online]. Available: \url{https://link.aps.org/doi/10.1103/PhysRevA.83.020302}
\BIBentrySTDinterwordspacing

\bibitem{kurpiers2018}
P.~Kurpiers, P.~Magnard, T.~Walter, B.~Royer, M.~Pechal, J.~Heinsoo, Y.~Salath{\'e}, A.~Akin, S.~Storz, J.-C. Besse \emph{et~al.}, ``Deterministic quantum state transfer and remote entanglement using microwave photons,'' \emph{Nature}, vol. 558, no. 7709, pp. 264--267, 2018.

\bibitem{axline2018}
C.~J. Axline, L.~D. Burkhart, W.~Pfaff, M.~Zhang, K.~Chou, P.~Campagne-Ibarcq, P.~Reinhold, L.~Frunzio, S.~Girvin, L.~Jiang \emph{et~al.}, ``On-demand quantum state transfer and entanglement between remote microwave cavity memories,'' \emph{Nature Physics}, vol.~14, no.~7, pp. 705--710, 2018.

\bibitem{azuma2023}
\BIBentryALTinterwordspacing
K.~Azuma, S.~E. Economou, D.~Elkouss, P.~Hilaire, L.~Jiang, H.-K. Lo, and I.~Tzitrin, ``Quantum repeaters: From quantum networks to the quantum internet,'' \emph{Rev. Mod. Phys.}, vol.~95, p. 045006, Dec 2023. [Online]. Available: \url{https://link.aps.org/doi/10.1103/RevModPhys.95.045006}
\BIBentrySTDinterwordspacing

\bibitem{li2023}
\BIBentryALTinterwordspacing
A.~Li, S.~Stein, S.~Krishnamoorthy, and J.~Ang, ``{QASMB}ench: A low-level quantum benchmark suite for {NISQ} evaluation and simulation,'' \emph{ACM Transactions on Quantum Computing}, vol.~4, no.~2, Feb. 2023. [Online]. Available: \url{https://doi.org/10.1145/3550488}
\BIBentrySTDinterwordspacing

\bibitem{bennett1996}
\BIBentryALTinterwordspacing
C.~H. Bennett, G.~Brassard, S.~Popescu, B.~Schumacher, J.~A. Smolin, and W.~K. Wootters, ``Purification of noisy entanglement and faithful teleportation via noisy channels,'' \emph{Phys. Rev. Lett.}, vol.~76, pp. 722--725, Jan 1996. [Online]. Available: \url{https://link.aps.org/doi/10.1103/PhysRevLett.76.722}
\BIBentrySTDinterwordspacing

\end{thebibliography}
